\documentclass[aps,reprint,amsmath,amssymb,groupedaddress]{revtex4-1}
\usepackage{graphicx}% Include figure files
\usepackage{dcolumn}% Align table columns on decimal point
\usepackage{bm}% bold math
\usepackage{color}
\begin{document}
%\bibliographystyle{apsrev4-1}

% The following information is for internal review, please remove them for submission
\widetext
%\leftline{Ver.4, submitted}
%\leftline{To be submitted to PRB}

% Use the \preprint command to place your local institutional report
% number in the upper righthand corner of the title page in preprint mode.
% Multiple \preprint commands are allowed.
% Use the 'preprintnumbers' class option to override journal defaults
% to display numbers if necessary
%\preprint{}

%Title of paper
\title{Planar tunneling spectroscopy of the topological Kondo insulator SmB$_6$}

% repeat the \author .. \affiliation  etc. as needed
% \email, \thanks, \homepage, \altaffiliation all apply to the current
% author. Explanatory text should go in the []'s, actual e-mail
% address or url should go in the {}'s for \email and \homepage.
% Please use the appropriate macro foreach each type of information

% \affiliation command applies to all authors since the last
% \affiliation command. The \affiliation command should follow the
% other information
% \affiliation can be followed by \email, \homepage, \thanks as well.
%\author{}
%\email[]{Your e-mail address}
%\homepage[]{Your web page}
%\thanks{}
%\altaffiliation{}
%\affiliation{}

\author{L. Sun,$^1$ D.-J. Kim,$^2$ Z. Fisk,$^2$ {and W. K. Park$^{1,\dag,}$}}
%\author{D.-J. Kim$^2$}
%\author{Z. Fisk$^2$}
%\author{W. K. Park$^{1,\dag}$}
 \email{wkpark@magnet.fsu.edu}
 \thanks{{\dag}Present address: National High Magnetic Field Laboratory, Florida State University, Tallahassee, Florida 32310, USA}
\affiliation{
$^1$Department of Physics and Materials Research Laboratory, University of Illinois at Urbana-Champaign, Urbana, Illinois 61801, USA\\
$^2$Department of Physics and Astronomy, University of California, Irvine, California 92697, USA
}

\date{\today}
%Collaboration name if desired (requires use of superscriptaddress
%option in \documentclass). \noaffiliation is required (may also be
%used with the \author command).
%\collaboration can be followed by \email, \homepage, \thanks as well.
%\collaboration{}
%\noaffiliation

\begin{abstract}
Several technical issues and challenges are identified and investigated for the planar tunneling spectroscopy of the topological Kondo insulator SmB$_6$. Contrasting behaviors of the tunnel junctions prepared in two different ways are analyzed and explained in detail. The conventional approach based on an AlO$_\text{x}$ tunnel barrier results in unsatisfactory results due to the inter-diffusion between SmB$_6$ and deposited Al. On the contrary, plasma oxidation of SmB$_6$ crystals produces high-quality tunnel barriers on both (001) and (011) surfaces. Resultant conductance spectra are highly reproducible with clear signatures for the predicted surface Dirac fermions and the bulk hybridization gap as well. The surface states are identified to reside on two or one {\it distinguishable} Dirac cone(s) on the (001) and (011) surface, respectively, in good agreement with the recent literature. However, their topological protection is found to be limited within the low energy region due to their inevitable interaction with the bulk excitations, called spin excitons, consistent with a recent theoretical prediction. Implications of our findings on other physical properties in SmB$_6$ and also other correlated topological materials are remarked.
\end{abstract}

% insert suggested PACS numbers in braces on next line
% insert suggested keywords - APS authors don't need to do this
%\keywords{}

%\maketitle must follow title, authors, abstract, \pacs, and \keywords
\maketitle
% body of paper here - Use proper section commands
% References should be done using the \cite, \ref, and \label commands
\section{Introduction\label{introduction}}
% Put \label in argument of \section for cross-referencing
%\section{\label{}}
%\subsubsection{}

The conventional Landau-Ginzburg paradigm based on the symmetries broken in ordered states breaks down in many topological phases of matter discovered recently \cite{Bernevig06,Konig07,Hasan10,Qi11}. Three-dimensional (3D) topological insulators (TIs) comprise one such class of emergent quantum matter, in which the nontrivial topology in the bulk band structure leads to topologically protected metallic states at surfaces \cite{Fu07,Hasan11}. Several dozens of 3D TIs have been discovered so far, among which Bi-based materials \cite{Fu07b,Zhang09}, such as Bi$_2$Se$_3$ \cite{Xia09,Hsieh09}, are best known. It is noteworthy that electron correlations do not play an important role in most of these materials. In recent years, Kondo insulators (KIs) \cite{Riseborough00} have drawn much attention due to a possibility that they might also be topological \cite{Dzero10}. Because strong correlations are at the heart of their underlying bulk physics, the surface states in these topological Kondo insulators (TKIs) are expected to exhibit more intriguing behaviors than in weakly correlated counterparts.

SmB$_6$, known as a Kondo or intermediate valence insulator (or semiconductor), has long been studied \cite{Menth69}. There is no doubt about the formation of a bulk hybridization gap below certain temperature and the appearance of metallic states at low temperature as first hinted by the resistivity plateau \cite{Allen79}. There had been several scenarios proposed to explain this exotic behavior, including impurity states in the bulk \cite{Gorshunov99}. However, the robustness of the plateau suggests that these conjectures are unlikely as such states should be easily destroyed by disorders. Coming at the right time, the theoretical proposal that certain Kondo insulators might be topological \cite{Dzero10}, followed by subsequent band structure calculations \cite{Takimoto11,Lu13,Kim14}, has been given a particular attention because it could readily explain the resistivity saturation behavior. These $f$-electron materials have inherently large spin-orbit coupling and the hybridization gap gets smaller with increasing correlation strength, fulfilling the requirement for TIs, i.e., band inversion \cite{Hasan11}. It was later elaborated that the cubic crystal structure \cite{Alexandrov13} and intermediate valence nature \cite{Dzero12} of SmB$_6$ would make it a prime candidate for TKI. Consequently, a resurgence of research interest has resulted in many new findings. By all means, including transport measurements \cite{Wolgast13,KimDJ14,Wakeham15,Luo15,Syers15}, it is now well established that the resistivity saturation is due to the robust metallic states at surfaces \cite{Dzero16,Allen16}. 

However, their detailed topological nature still remains to be unraveled unlike in the case of weakly correlated counterparts, e.g., Bi$_2$Se$_3$, in which various measurements, including angle- and spin-resolved photoemission spectroscopy (ARPES) \cite{Xia09,Hsieh09} and scanning tunneling spectroscopy (STS) \cite{Cheng10,Hanaguri10}, have shown that the surface Dirac fermions are indeed topological with the predicted spin-momentum locking nature. Although several photoemission measurements on SmB$_6$ \cite{Miyazaki12,Denlinger13,Neupane13,Jiang13,Frantzeskakis13,Min14,Xu14} have revealed the formation of a hybridization gap below $\sim$50 K and in-gap states, the exact topological nature is not unveiled yet. Also, clear signatures for the surface Dirac fermions such as linear conductance are lacking in several STS \cite{Yee13,Rossler14,Ruan14} and point-contact spectroscopy \cite{Zhang13} measurements. The challenges encountered in studying SmB$_6$, or TKIs in general, are manifold. From the materials science point of view, these materials are much less favorable for those surface-sensitive spectroscopic measurements than the Bi-based materials. More specifically, while the layered structure of Bi$_2$Se$_3$ allows an easy exposure of atomically smooth surfaces via cleavage, the 3D crystalline structure of SmB$_6$ makes it unfeasible. Also, it is known that the polar nature of the (001) surface, where the topological surface states are predicted to exist, causes various complex issues including surface reconstruction \cite{Yee13} and time-dependent evolution \cite{Zhu13}. Quantum oscillation measurements have shown the existence of surface bands and their possible topological nature \cite{Li14} but the nature of the bulk insulating state inferred from such measurements is currently under debate \cite{Tan15}. From the fundamental physics point of view, the complexity has to do with the strong correlations in these materials underlying their topological origin.

We have addressed some of the aforementioned challenges and obtained new spectroscopic information on the topological nature in SmB$_6$ \cite{Park16}. Here, we adopt planar tunneling spectroscopy (PTS) \cite{Giaever61,McMillan65,Wolf85} because it is an inherently surface-sensitive probe \cite{Covington97}. In addition, the narrow tunneling cone in PTS geometry \cite{Beuermann81} may allow momentum-selective measurements, which are useful when it is necessary to distinguish signals originating from different Dirac cones, as seen in our work. In contrast, despite its clear advantage of space-resolved spectroscopic mapping capability \cite{Fischer07}, STS may not allow such measurements due to an inherently much wider tunneling cone. There is another point to make in comparing spectroscopic techniques: in PTS (also in STS), a complete energy range, i.e., both below and above the chemical potential, can be probed easily by reversing the bias polarity, whereas in ARPES it is non-trivial to obtain signals above the chemical potential. This difference clearly stands out in our tunneling study \cite{Park16}. 

A basic underlying principle for PTS is Fermi's golden rule. A simple tunnel junction is comprised of a bottom electrode (typically, the material of interest), a tunnel barrier, and a top (or counter) electrode with a constant density of states (DOS) near the Fermi level ($N_\textrm{c}(0)$), that is, a simple metal. Then, the differential conductance, $G$(V) $\equiv$ $dI/dV$, is simply given by a convolution of the DOS of the material of interest ($N_\textrm{s}(E)$) with respect to the derivative of the Fermi function ($f(E)$):   
\begin{equation}
\frac{dI}{dV} = A |M|^2 e^2N_c(0)\int\limits_{-\infty}^{\infty}N_s(E)\frac{\partial f(E-eV)}{\partial (eV)}dE,
\end{equation}
where $A$ is the junction area and $M$ is the tunneling matrix element. Thus, a measurement of tunnel conductance can reveal detailed DOS structures.

For high quality, the tunnel barrier should be sharply interfaced with both electrodes and its electrostatic potential should be much higher than the maximum bias voltage. Typically, it is made of a thin layer of insulating oxide such as AlO$_\text{x}$ \cite{Giaever61,Wolf85}. Depending on the constituent materials, depositing or forming such a thin barrier layer may involve some challenges. This is particularly the case with SmB$_6$, for which the conventional AlO$_\text{x}$ barrier is found quite unfavorable. On the contrary, plasma oxidation of the crystal surface works quite well to turn the top surface into a tunnel barrier. Combined with other cooperative factors including the excellent polishability of SmB$_6$ crystals and the use of superconducting Pb as a counter-electrode, this approach allows us to obtain reproducible conductance spectra.

In this paper, we shall focus on experimental developments around the two approaches since detailed analysis and interpretation of the reproducible conductance features have been reported elsewhere \cite{Park16}. In the next section, experimental details regarding the junction fabrication and characterization are described. In Sect. III, results from the above-mentioned two different approaches are presented. Those from initial attempts and also from later systematic diagnostic runs using an AlO$_\text{x}$ barrier are reported first, followed by the description of the data from junctions made of plasma-oxidized SmB$_6$ surface. Sect. IV contains detailed discussion of the failures and successes observed in these approaches as well as a brief discussion on the topological nature in SmB$_6$. A summary follows in the last section.

\section{Experiments}

\subsection{Crystal Surface Preparation}

Frequently, for PTS, the sample is in a thin-film form onto which a uniform insulating barrier can be easily deposited. Here, we chose to use high-quality single crystals grown by a flux method \cite{KimDJ14}. Since the surface of a typical as-grown SmB$_6$ crystal is not so smooth as desired for tunnel junction fabrication, it has to be polished to mirror-like shininess. For this, as-grown crystals with lateral dimensions of 1 -- 2 mm and thickness of $\sim$0.5 mm are first embedded into molds made of low-temperature epoxy (Stycast\textsuperscript{\textregistered}, 2850-FT). 

\begin{figure}[tbp]
\centering
\includegraphics[keepaspectratio, width=0.9 \columnwidth]{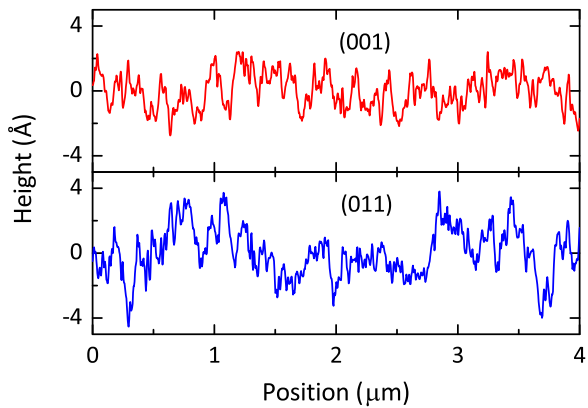}
\caption{Linear topographic profiles of the polished (001) and (011) surfaces of SmB$_6$ crystals measured with an AFM, showing that the polished surfaces are smooth enough for the deposition/formation of a thin tunnel barrier.}
\label{AFM}
\end{figure}

Polishing is done mechanically using alumina lapping films of particle sizes ranging from 12 -- 0.3 $\mu$m. The crystal is pressed manually with moderate pressure and rubbed against the lapping film. Isopropyl alcohol is sprayed whenever lubrication is necessary. The polishing lasts for about 10 -- 15 minutes using each lapping film, and the surface is subsequently inspected under an optical microscope. If it appears to be smoother and more uniform, the polishing proceeds with a lapping film of smaller particle. Figure \ref{AFM} displays cross-sectional topographic profiles obtained with an atomic force microscope (AFM).

%[keepaspectratio, width=0.9 \columnwidth]

\subsection{Junction Fabrication}

The polished crystal is loaded into a high-vacuum chamber. First, an Ar ion beam is used to  clean the surface by etching out, if any, surface oxides or contaminants residual from the polishing process. We have adopted two methods to form a tunnel barrier: (i) Deposition and subsequent oxidation of a thin Al film to form AlO$_\text{x}$; (ii) Plasma oxidation of the crystal's top surface. In the former approach, the Al film is deposited by dc magnetron sputtering at room temperature unless otherwise specified. The deposition rate ranges from 1.25 -- 5.0 {\AA} s$^{-1}$. Al films of various thicknesses, $d_\textrm{Al}$, between 15 and 70 {\AA} have been used to optimize the parameters in either plasma \cite{Kuiper01} or thermal oxidation process. The oxygen plasma is generated by dc glow discharge. For thermal oxidation, the crystal is left in the vacuum chamber filled with 1 -- 10 Torr of oxygen for half an hour. In the second method, the plasma oxidation is done similarly to the first case but without the Al layer.

\begin{figure}[tbp]
\centering
\includegraphics[keepaspectratio, width=0.9 \columnwidth]{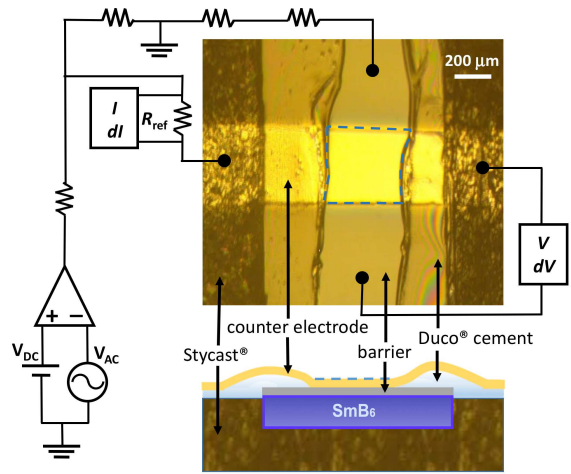}
\caption{The structure of a typical tunnel junction on SmB$_6$. Top panel is an optical image of a real junction and the bottom panel is its schematic cross-sectional diagram. The dashed rectangle indicates the junction area. Also shown is the wiring configuration for conductance measurements using a custom-built mixing circuit.}
\label{jcn-structure}
\end{figure}

Prior to the deposition of counter-electrodes, crystal edges are painted with diluted cement in order to prevent them from being shorted to the bottom electrode (SmB$_6$) and also to ensure that junctions could be defined over the most uniform area (see Fig. \ref{jcn-structure}). We adopt Duco\textsuperscript{\textregistered} cement because of its confirmed stability over thermal cycling. Once the cement is cured, counter electrodes are thermally evaporated through a shadow mask that is carefully aligned under a microscope. We use Pb (Ag) as a (non-)superconducting counterelectrode since it is easy to evaporate thermally. The deposition is done at a moderate rate of 8 -- 10 {\AA} s$^{-1}$ to prevent damages to the barrier layer and the total thickness is typically 2500 {\AA}. Owing to its sharp superconducting DOS, Pb is found to serve as an excellent filter for the quality of junctions. The detailed junction structure is displayed in Fig. \ref{jcn-structure}, which also shows the wiring configuration for our conductance measurements.

\subsection{Junction Characterization}

There are several ways to check the quality of junctions. First, if they are made with well-defined structures including a uniform tunnel barrier, their differential resistance ($R_\textrm{J} \equiv dV/dI$) should be inversely proportional to their area and, in turn, the $R_\textrm{J}A$ product would be nearly constant. Our optimized SmB$_6$ junctions typically have $R_\textrm{J}A$ values in the range of 50 -- 150 $\Omega$mm$^2$. Those with too large $R_\textrm{J}A$ usually show large fluctuations as the bias voltage is ramped and those with too small $R_\textrm{J}A$ tend to have leakage currents due to micro-shorts across the barrier. If the $R_\textrm{J}A$ value at room temperature is found to be in the favorable range, we proceed by recording the zero-bias conductance (ZBC) as a function of temperature.  For a material which experiences noticeable changes in its electronic states like the gap opening in SmB$_6$, the temperature dependence of ZBC is to reflect such changes inevitably. Once the temperature is stabilized at the base, $G(V)$ is measured over wide ranges of temperature and magnetic field. When Pb is used as a counterelectrode, if necessary, a magnetic field of 0.1 T is applied to suppress its superconductivity.

\section{Results}

\subsection{SmB$_6$/AlO$_\text{x}$/Ag(Pb) junctions}
 
Although an AlO$_\text{x}$ barrier has been widely adopted for PTS \cite{Giaever61,Wolf85}, realizing it on a specific material of interest could be non-trivial. The resulting junction might suffer from under- or over-oxidation. Slight under-oxidation would hamper detecting the intrinsic DOS albeit less serious in superconductive junctions owing to the proximity effect. Over-oxidation is also problematic because it could form additional oxides originating from the material itself. Thus, in order to produce high-quality tunnel junctions, the oxidation process should be optimized such that the Al layer is completely oxidized while the SmB$_6$ crystal remains intact.

\begin{figure}[tbp]
\centering
\includegraphics[keepaspectratio, width=0.9 \columnwidth]{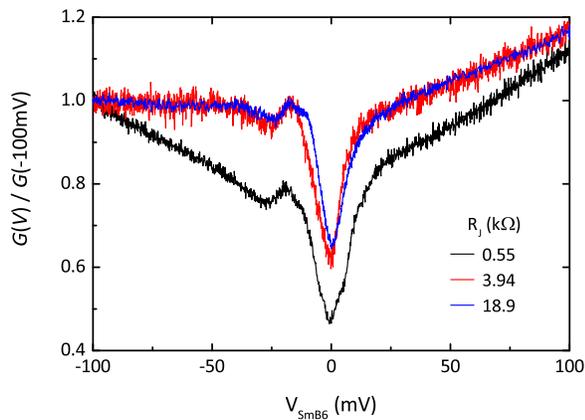}
\caption{Normalized $G(V)$ data taken at 1.7 K for three (001)SmB$_6$/AlO$_\text{x}$/Ag junctions showing some features due to SmB$_6$. The $R_\textrm{J}$ values are given at $-100$ mV. $d_\textrm{Al}$ (oxidation conditions) is 50 (plasma, 4.2 W for 30 seconds), 50 (thermal, 1 Torr for 30 minutes), and 20 {\AA} (plasma, 1.94 W for 20 seconds), in the order of increasing $R_\textrm{J}$.
%The conductance suppression around zero bias and the peak around $-$(17 $\sim$ 20) mV indicate the opening of a hybridization gap. The overall asymmetric shape is due to a Fano interference effect. A faint signature for the surface Dirac fermions is also observed as the linearity at low bias, in particular, for the junction with $R_\textrm{J}$ = 0.55 k$\Omega$.
}
\label{BestAlOx}
\end{figure}

More than fifty junction fabrication cycles have been executed repeatedly using two crystals with (001) or (011) surface orientation, respectively. Only a very small portion (less than 4\%) of the junctions show reproducible conductance features, as displayed in Fig. \ref{BestAlOx}. They commonly exhibit both a gap-like suppression around zero bias and an overall asymmetric shape. While a broad peak is seen in the negative bias branch, around $-$(17 $\sim$ 20) mV, the conductance increases monotonically in the positive bias counterpart. As this asymmetry can be attributed to a Fano interference effect \cite{Fano61} in a Kondo lattice (or Anderson lattice, more broadly) \cite{Maltseva09,Figgins10,Wolfle10}, the conductance data must reflect intrinsic DOS in SmB$_6$ to some extent. The conductance curves appear to be linear at low bias (in particular, see the curve for $R_\textrm{J}$ = 0.55 k$\Omega$), suggesting the existence of surface Dirac fermions that have V-shaped DOS, but detailed features are buried due to large fluctuations. The overall features observed in these curves are qualitatively similar to those revealed in the best-quality junctions prepared by surface oxidation \cite{Park16} (also, see Fig. \ref{SmB6reprod}), presumably because the adopted processing parameters happen to allow the formation of rather a clean tunnel barrier. However, due to the poor reproducibility, in the following we shall focus on understanding how different parameters might affect the formation and quality of an AlO$_\text{x}$ barrier on SmB$_6$.  

\begin{figure}[tbp]
\centering
\includegraphics[keepaspectratio, width=0.9 \columnwidth]{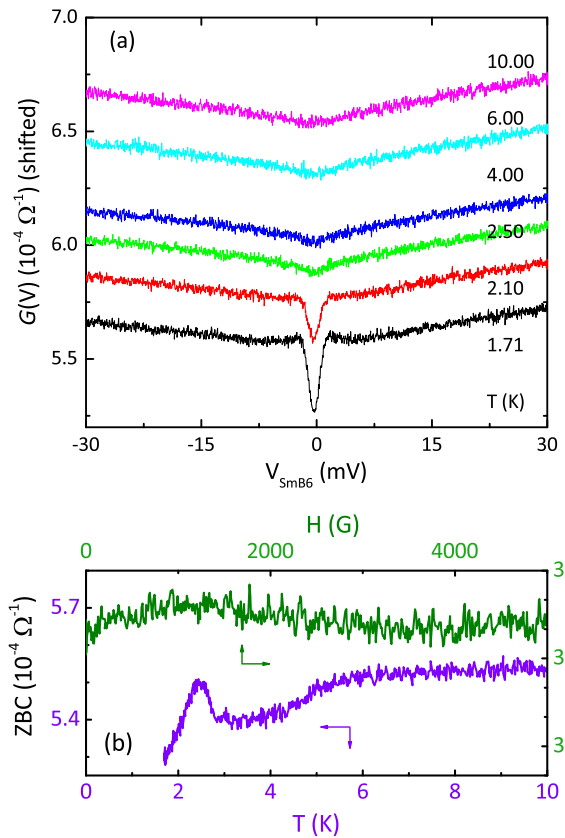}
\caption{Conductance of a (001)SmB$_6$/AlO$_\text{x}$/Ag junction dominated by the superconductivity in unoxidized Al. $d_\textrm{Al}$ = 30 {\AA} and the barrier is formed by thermal oxidation. (a) Conductance spectra as a function of temperature. For clarity, the curves are shifted vertically. (b) Temperature- (bottom axis) or magnetic-field- (top axis) dependent ZBC.}
\label{UnoxidizedAl}
\end{figure}

\begin{figure}[tbp]
\centering
\includegraphics[keepaspectratio, width=0.9 \columnwidth]{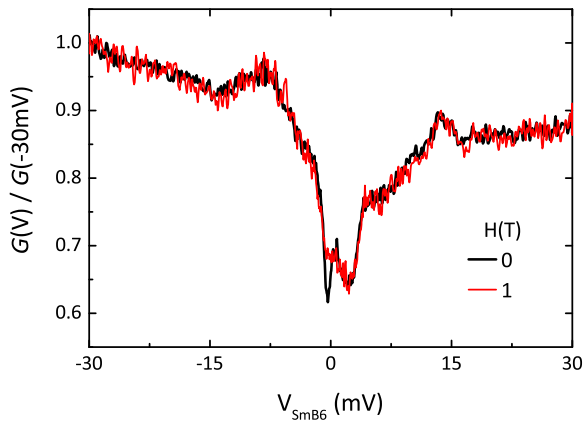}
\caption{Normalized conductance of a (011)SmB$_6$/AlO$_\text{x}$/Ag junction at 1.37 K, exhibiting signatures due to both SmB$_6$ and unoxidized Al. $d_\textrm{Al}$ = 40 {\AA} and the barrier is formed by thermal oxidation.}
\label{mixedcond}
\end{figure}

Junctions with an AlO$_\text{x}$ barrier frequently have too large resistance ($R_\textrm{J}A > 1000$ $\Omega$mm$^2$). This causes large conductance fluctuations, rendering detailed DOS features buried in noise. One may consider decreasing the barrier thickness as a solution but what actually happens is quite complicated as detailed later. Figure \ref{UnoxidizedAl} shows an example in which the conductance does not reveal any DOS features of SmB$_6$. While the conductance is only semi-linear at high temperatures, a gaplike suppression appears with decreasing temperature. The peaky shape around $\pm$2 mV is reminiscent of superconducting coherence peaks, which are unexpected if the entire Al layer were oxidized. The ZBC data as a function of temperature and magnetic field, shown in Fig. \ref{UnoxidizedAl}(b), indicate that the ZBC suppression disappears above $\sim$2.3 K and $\sim$1000 G. This suggests that indeed the gap-like features might originate from superconductivity. A plausible speculation is that the bottom part of the Al layer might be left unoxidized, prohibiting the tunneling electrons from feeling the DOS in SmB$_6$ directly. Then, why is its $T_\textrm{c}$ higher than in the bulk Al (1.2 K)? This can be understood by considering that the unoxidized Al layer might be disordered, in which case the $T_\textrm{c}$ can go up \cite{Dynes81}. This observation may appear to point to the importance of complete oxidation of the Al layer. However, it turns out more complicated than it appears. For instance, the conductance spectra from some other junctions reveal more complex features as plotted in Fig. \ref{mixedcond}. Here, the conductance is asymmetric with some gap-like features including the broad peak centered around a negative bias and the suppression around zero bias, whose origin might be similar as in Fig. \ref{BestAlOx}. But there also appear additional features at low bias ($\pm$1 mV). From their suppression under a magnetic field, reminiscent of the behavior seen in Fig. \ref{UnoxidizedAl}(b), again we associate them with superconductivity in the unoxidized Al. To understand how those mixed features can appear, we speculate that, under some unknown conditions, the oxidation may result in a nonuniform barrier with the Al being almost completely oxidized over some areas but remaining unoxidized in other areas.

\begin{figure*}[tbp]
\centering
\includegraphics[keepaspectratio, width=1.8 \columnwidth]{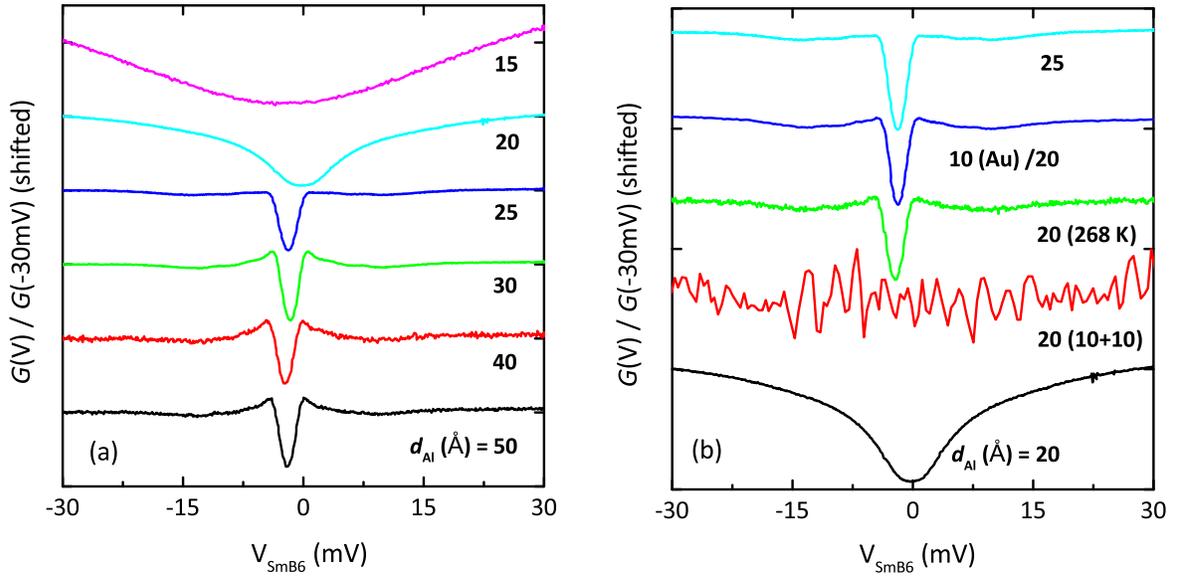}
\caption{Normalized conductance spectra of SmB$_6$/AlO$_\text{x}$/Pb junctions prepared by thermal oxidation with varying $d_\textrm{Al}$. The curves are taken at 4.2 K and shifted vertically for clarity. The SmB$_6$ crystal has either (001) or (011) orientation, which is not specified here since the results do not seem to depend on it. (a) Evolution of normalized $G$(V) with $d_\textrm{Al}$ decreasing from 50 to 15 {\AA}. (b) Conductance spectra of junctions with nominally the same Al thickness (20 {\AA}), revealing the impact of how the Al layer is deposited (three middle curves). The curves for $d_\textrm{Al}$ = 20 and 25 {\AA} from (a) are also included for comparison.}  
\label{interdiffusion}
\end{figure*}

In order to investigate whether optimizing the oxidation parameters is a major issue, we performed experiments for a series of junctions with $d_\textrm{Al}$ decreasing from 50 {\AA} to 15 {\AA} and Pb as the counter-electrode. Here, we adopt thermal oxidation instead of plasma oxidation for the consistency of oxidation parameters (1 Torr, 30 min.) among different runs. If the oxidation itself were the issue, intrinsic DOS features could be observed reproducibly once $d_\textrm{Al}$ is optimized. Figure \ref{interdiffusion}(a) shows their normalized $G(V)$ curves at 4.2 K. Junctions with thicker Al exhibit the Pb coherence peaks but no features due to SmB$_6$, which is possible again if part of the Al layer is left unoxidized. The Pb gap features are suppressed quite slowly with decreasing $d_\textrm{Al}$ down to 25 {\AA}, below which they disappear abruptly and only a broad dip develops around zero bias. Combined with results from many other runs, we conclude that junctions with a very thin Al ($d_\textrm{Al} <$ 15 {\AA}) do not show any SmB$_6$ features at all regardless of how it is oxidized. This observation is contradictory to our initial speculation mentioned above.

One might suspect the uniformity of the Al layer deposited. To address this issue, we employed Auger electron spectroscopy to measure the chemical homogeneity from several different spots on the surface of SmB$_6$ single crystals coated with 20-{\AA}-thick Al. The statistics clearly shows that the relative intensity of the Al peak is uniform on both (001) and (011) surfaces, ruling out that possibility. Also, the fact that the superconducting Pb features are still observed for $d_\textrm{Al}$ = 25 {\AA} implies that the unoxidized Al is quite uniform over the junction area. This reasoning led us to speculate that the culprit might be at the interface between SmB$_6$ and Al.

To investigate this possibility, three more experiments are conducted by preparing junctions with nominally the same $d_\textrm{Al}$ (= 20 {\AA}) but deposited in three different ways as follows. In one batch, 10 {\AA}-thick Al is deposited and oxidized thermally, which is repeated twice with an anticipation that the Al  layer could be oxidized more completely and uniformly. However, as shown in Fig. \ref{interdiffusion}(b), this method does not lead to any better data but featureless and noisy signals. We then speculate this might be because the resultant AlO$_\text{x}$ layer is highly disordered. Thus, the next run is carried out by depositing 20 {\AA}-thick Al under better base vacuum achieved by running a liquid nitrogen jacket inside the chamber. In this case, more residual moisture is expected to be removed so that the resultant AlO$_\text{x}$ barrier could be cleaner. Quite surprisingly, the superconducting Pb features reappear in this junction, contrary to the case  in Fig. \ref{interdiffusion}(a) with the same $d_\textrm{Al}$ but processed without the nitrogen jacket running. This implies that whatever causing the abrupt change in the conductance behavior going from 25 {\AA} to 20 {\AA} in Fig. \ref{interdiffusion}(a) disappears. Because the conductance features are almost identical with the 25 {\AA} case without showing any SmB$_6$ features anticipated for thinner Al, this behavior can not be explained by the cleanliness of the barrier itself. In this regard, we note that the sample stage temperature during the deposition dropped to $\sim$268 K due to convective cooling by the running nitrogen jacket. This raises a possibility that Al atoms might diffuse into SmB$_6$ (or vice versa) at room temperature, particularly in the initial stage, whereas such diffusion is greatly reduced at low temperature to allow them to form a sharper interface with SmB$_6$. In turn, if the Al layer were not oxidized completely, the superconducting Pb features would reappear. To test this scenario in another way, the last trial is to deposit a thin (10 {\AA}) Au layer prior to Al. This is a feasible scheme since different elemental atoms would have much different diffusion constants. Here, both Au and Al layers are deposited at room temperature as usual. If the inter-diffusion were a real issue, this thin Au layer might alleviate it substantially by acting as a diffusion barrier.  Indeed, the Pb superconducting features reappear as shown in Fig. \ref{interdiffusion}(b), very similarly to the previous one. These observations indicate that junctions having nominally the same $d_\textrm{Al}$ behave very differently depending on how the Al layer is deposited due to the inter-diffusion between Al and SmB$_6$.

\subsection{SmB$_6$/Oxi-SmB$_6$/Pb junctions}

\begin{figure*}[tbp]
\begin{center}
\includegraphics[keepaspectratio, width= 1.8  \columnwidth]{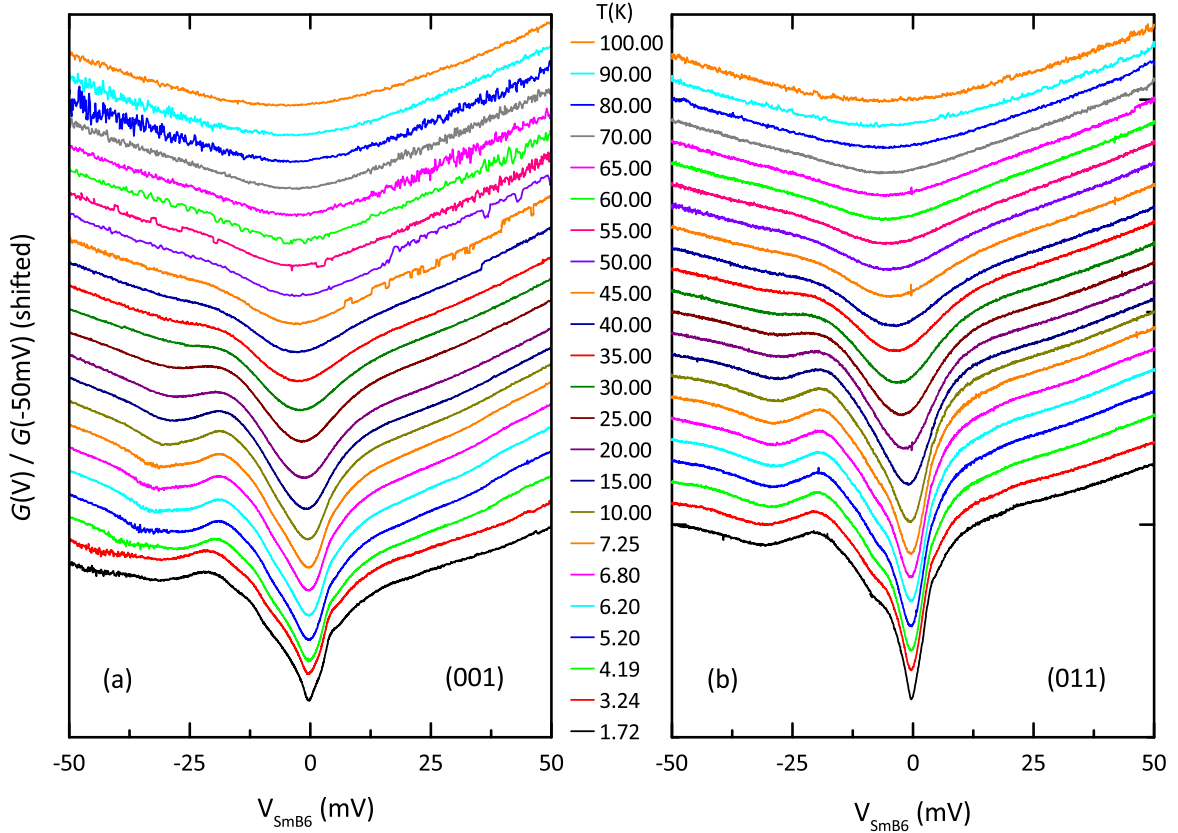}
\end{center}
\caption{Normalized $G$(V) of SmB$_6$/Oxi-SmB$_6$/Pb junctions for (a) (001) and (b) (011) surface orientations. The barrier is prepared by plasma oxidation of the SmB$_6$ crystal. The data below $T_\textrm{c}$ (7.2 K) of Pb were taken with the superconductivity suppressed by an applied magnetic field of 1000 G. The curves are shifted vertically for clarity.}
\label{SmB6cond}
\end{figure*} 

As an alternative to the traditional AlO$_\text{x}$ barrier, we have tried several other methods, including deposition of a thin Nb layer as a diffusion barrier and/or buffer layer for Al or oxidation of that layer to form a tunnel barrier. While some of these trials give better results, high-quality junctions are not obtained reproducibly by any approach except plasma oxidation of the SmB$_6$ crystal surface \cite{Park16}. These tunnel junctions, denoted as SmB$_6$/Oxi-SmB$_6$/Pb, not only exhibit intrinsic DOS features due to SmB$_6$, as shown in Fig. \ref{SmB6cond}, but also highly reproducible. 

\begin{figure*}[tbp]
\centering
\includegraphics[keepaspectratio, width=1.75 \columnwidth]{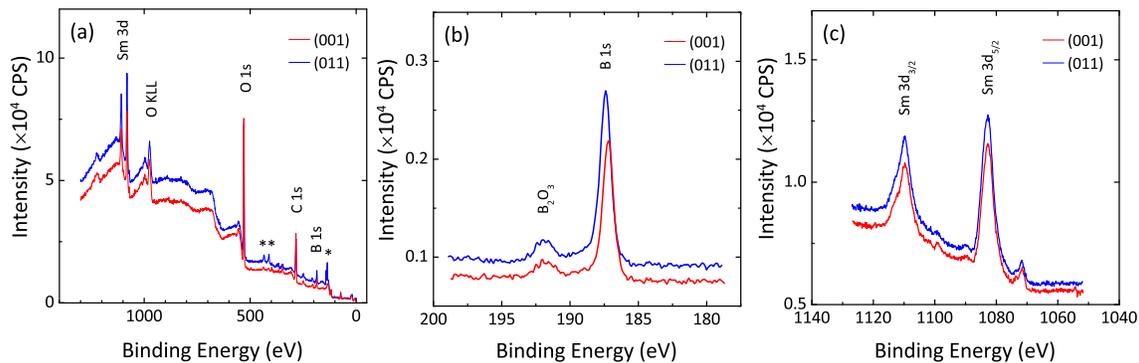}
\caption{XPS data from plasma-oxidized SmB$_6$ crystal surfaces. (a) Spectra over the entire binding energy range. The peaks labeled with asterisks are due to the epoxy mold. (b) Detailed data to show the chemical states of B atoms. (c) Peaks arising from Sm 3$d$ electrons.}
\label{xps}
\end{figure*} 

The temperature dependence displayed in Fig. \ref{SmB6cond} provides a detailed picture on how the electronic states in SmB$_6$ evolve from a bad metallic behavior at high temperature to an insulating (or semiconducting) to a surface-conduction dominant state at low temperature. The bulk hybridization gap, as evidenced by the broad peak around $-$21 mV as well as the low-bias conductance suppression, appears to form below $\sim$50 K but signatures for the surface states do not stand out until the temperature is lowered further down to $\sim$25 K, as seen more clearly in $G(V_b, T)$ curves \cite{Park16}. Afterward, the surface states appear to undergo quite a complicated multi-step evolution, which we associate with their interaction with the bulk excitations, called spin excitons \cite{Fuhrman15}.

In order to investigate what kind of tunnel barrier is formed in these junctions, we have carried out x-ray photoelectron spectroscopy (XPS) measurements on the SmB$_6$ crystals oxidized similarly. Figure \ref{xps} shows XPS data on both (001) and (011) surfaces. In Fig. \ref{xps}(b), some of the boron atoms are found to be oxidized to form B$_2$O$_3$ \cite{Ong04}. On the other hand, no clear evidence is found for oxidized Sm atoms as shown in Fig. \ref{xps}(c). While the formation of some samarium oxides such as Sm$_2$O$_3$ can not be completely ruled out due to their chemical complexity \cite{Cho07}, we think that the B$_2$O$_3$ layer serves as a tunnel barrier, as also supported by a band structure calculation \cite{Li96} showing that B$_2$O$_3$ has a large (6--9 eV) band gap.

Also shown in Fig. 7, SmB$_6$/Oxi-SmB$_6$/Pb junctions exhibit linear conductance shape at low bias reminiscent of the V-shaped DOS of Dirac fermions, which is distinctly different between the two surfaces, i.e., double- versus single-linear. As has been explained in detail in Ref. \cite{Park16}, this is interpreted as due to the difference in the number of distinct Dirac cones, which is detected presumably due to PTS' momentum-selective nature \cite{Beuermann81}. Our observation is in good agreement with both theoretical calculations \cite{Takimoto11,Lu13,Kim14,Alexandrov13} and quantum oscillation measurements \cite{Li14}. 

Another significant spectroscopic feature is the linearity ending well below the hybridization gap edges. As described in Ref. \cite{Park16}, the surface states in SmB$_6$ are prone to the interaction with bulk excitations, spin excitons, abundant due to its close proximity to an antiferromagnetic quantum critical point \cite{Kapilevich15} as detected in recent neutron scattering measurements \cite{Fuhrman15}. Our conductance spectra clearly exhibit the features evidencing such interaction, also seen in some ARPES measurements \cite{Neupane13,Arab16}.

\begin{figure}[tbp]%[h]
\centering
\includegraphics[keepaspectratio, width=0.85 \columnwidth]{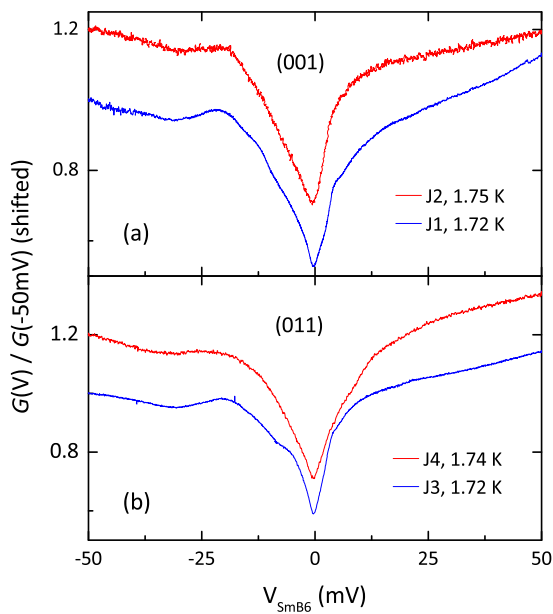}
\caption{Reproducibility of conductance spectra in SmB$_6$/Oxi-SmB$_6$/Pb junctions. Normalized conductance curves shown for two junctions prepared in different runs, shifted vertically by 0.2 for clarity.}
\label{SmB6reprod}
\end{figure}

The features discussed above are quite reproducible. Figure \ref{SmB6reprod} shows conductance data obtained from junctions prepared in different runs. The bulk gap features including the broad peak at $-$21 mV and the suppression around zero bias are clearly observed. Also, features due to the topological surface states are detected, including the double- versus single-linear conductance shape and the kink-hump structure originating from their interaction with bulk excitations. The two junctions for a given surface orientation show slightly different Dirac points and kink locations as well, which might originate from the difference in chemical potential and also in length scale for the interaction depending on detailed conditions for the plasma oxidation.

%\vspace{20mm}

\section{Discussion}

In the last section, several issues and challenges are identified in producing high-quality tunnel junctions on SmB$_6$. A great number of junctions based on the conventional AlO$_\text{x}$ barrier exhibit unpredictable behaviors ranging from trivial conductance shape such as parabolic or semi-linear, superconducting features due to the under-oxidized Al layer, to features somewhat intrinsic to SmB$_6$. A few systematic studies with various $d_\textrm{Al}$ show that the culprit is the inter-diffusion between SmB$_6$ and Al. Here we address these issues in more detail by considering realistic junction structures and tunneling processes.

\begin{figure*}[tbp]
\centering
\includegraphics[keepaspectratio, width=1.7 \columnwidth]{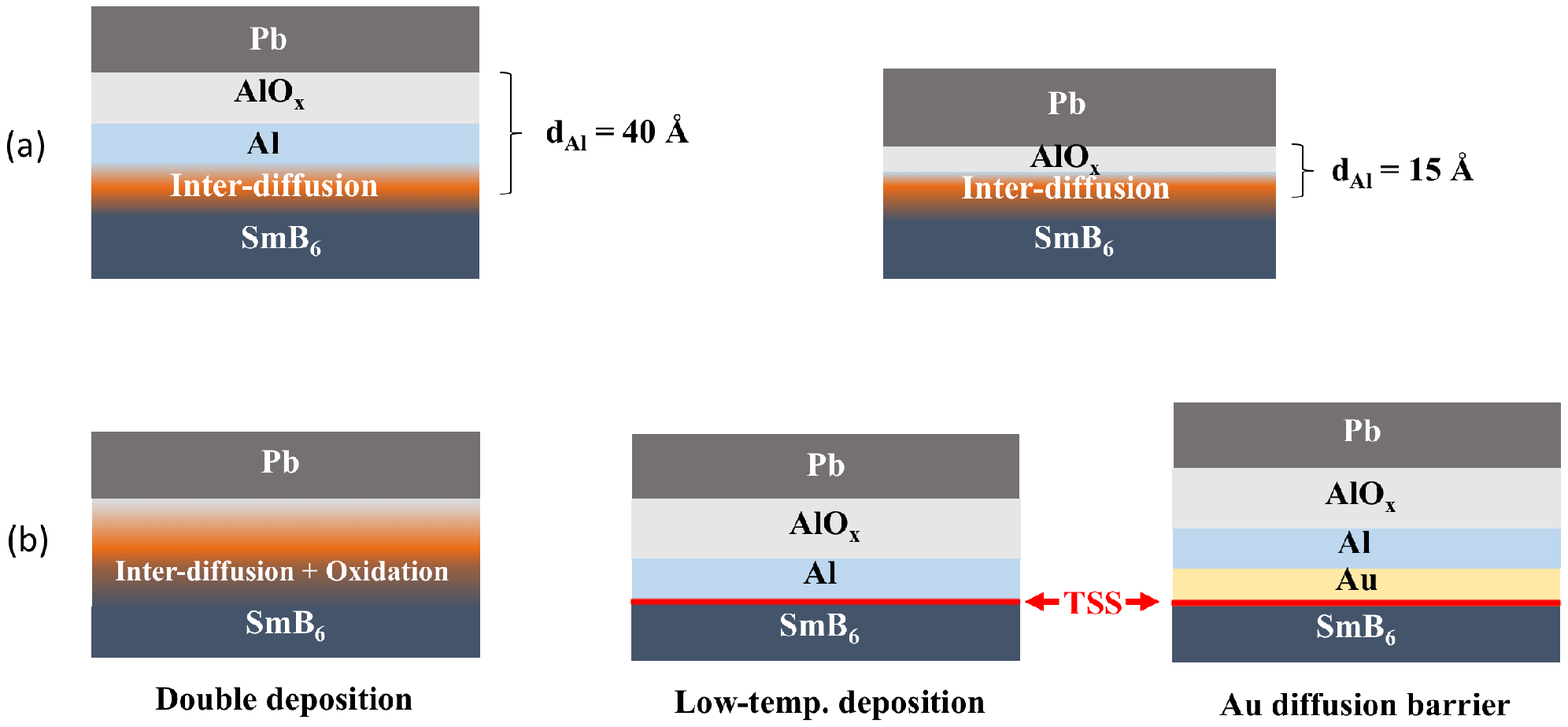}
\caption{Schematic cross-sectional diagrams of the SmB$_6$/AlO$_\text{x}$/Pb junctions prepared by thermal oxidation. (a) To illustrate contrasting behaviors between junctions with an Al layer in the thicker (40 {\AA}) or thinner (15 {\AA}) limit. The left panel is for $d_\textrm{Al} \geq$ 25 {\AA}, thick enough to leave a part of the Al layer unoxidized on top of the inter-diffusion layer. The right panel shows the opposite case, namely, $d_\textrm{Al}$ $<$ 25 {\AA}, where the Al layer is too thin to leave an unoxidized layer. (b) Structures of three junctions with nominally the same Al thickness (20 {\AA}) but deposited and oxidized in different ways. Left panel: 10 {\AA}-thick Al is deposited and oxidized, which is repeated twice. Middle panel: 20 {\AA} is deposited at low temperature ($\sim$268 K). Right panel: 10 {\AA}-thick Au is deposited prior to the Al layer at room temperature. TSS denotes the topological surface states.}
\label{jcnlayer}
\end{figure*}

Let us begin by discussing what might happen as $d_\textrm{Al}$ is varied. In Fig. \ref{jcnlayer}(a), we consider two limiting cases. When the Al layer is too thick, regardless of what actually happens at the interface, some Al will remain unoxidized underneath the top AlO$_\text{x}$ layer. Then, the tunneling will occur between the top electrode and the unoxidized Al instead of SmB$_6$, which is why some junctions exhibit features due to superconducting Al (Fig. \ref{UnoxidizedAl}) or Pb (Fig. \ref{interdiffusion}). When the Al layer is too thin, even if a good-quality AlO$_\text{x}$ barrier is formed at the top, due to the inter-diffusion layer underneath, electrons will go through diffusive transport instead of single-step elastic tunneling into SmB$_6$. This explains why the Pb superconducting features disappear abruptly as $d_\textrm{Al}$ is decreased from 25 {\AA} to 20 {\AA} (see Fig. \ref{interdiffusion}(a)).

Figure \ref{jcnlayer}(b) illustrates structures of the three junctions with nominally the same $d_\textrm{Al}$ (20 {\AA}), whose conductance spectra are plotted in Fig. \ref{interdiffusion}(b). First, depositing and oxidizing a thinner (10 {\AA}) Al twice to ensure a thorough oxidation does not work since the first Al deposition has already resulted in an inter-diffusion layer, whose oxidation would produce some non-conducting complex oxides. Thus, even if the second Al layer turned into a good tunnel barrier, the conductance data would not reveal any information on the electronic states in SmB$_6$ but featureless noise. Second, when the Al layer is deposited while the liquid nitrogen jacket is running, the lowered sample temperature reduces the inter-diffusion. Thus, most of the 20 {\AA}-thick Al layer remains intact, part of which is left unoxidized under the thermal oxidation conditions adopted, as depicted in the middle panel of Fig. \ref{jcnlayer}(b). This explains why the Pb superconducting features reappear. Lastly, the thin Au layer (10 {\AA}) prevents the inter-diffusion as it may act as a diffusion barrier, as illustrated in the right panel of Fig. \ref{jcnlayer}(b).   

The inter-diffusion layer may consist of some alloy-like structures in which the Al atoms randomly occupy the lattice sites in SmB$_6$. While the length scale over which such diffusion occurs remains to be investigated, it is clear that the lattice translational symmetry is broken in that region, preventing the formation of coherent heavy bands in a Kondo lattice. In turn, there will not be any topological surface states since, theoretically, they can arise only under the formation of coherent states in the bulk or a hybridization gap \cite{Dzero10,Dzero16}. The topological surface states might still exist beyond the inter-diffusion layer, similarly to the case of ion-damaged SmB$_6$ crystals \cite{Wakeham15}, but they would not be felt by tunneling electrons unlike in transport measurements, in which the current is set to flow along the surface. The reason why the SmB$_6$ features are not still observed when the inter-diffusion is reduced by depositing 20-{\AA}-thick Al at low temperature may have to do with the length scale relevant for tunneling electrons. We speculate that the unoxidized Al layer might be still too thick ($>$ 10 {\AA}) to allow tunneling electrons' wave functions to extend down to the surface state region. 

\begin{figure}[tbp]%[h]
\centering
\includegraphics[keepaspectratio, width=0.85 \columnwidth]{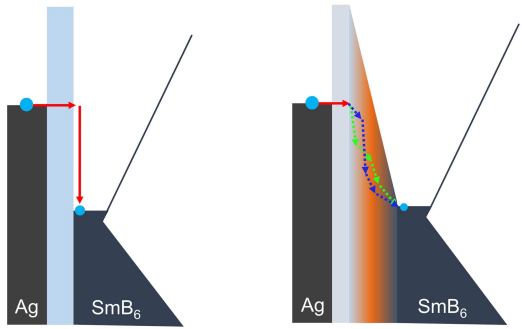}
\caption{Diagrams to illustrate the tunneling process in SmB$_6$ junctions. For simplicity, only transmissive (not reflective) trajectories are considered. Left panel: When the junction is of high quality with the barrier formed by plasma oxidation of the crystal, a single-step elastic tunneling is predominant. Right panel: When the barrier is formed by depositing and oxidizing a thin Al layer, the charge transport is dominated by multiple diffusive steps in the inter-diffusion layer as illustrated by the two different trajectories.}
\label{tunnelmodel}
\end{figure}

In order to enhance our microscopic understanding, we speculate on what trajectories the electrons will follow when forced to move from the counter-electrode (Ag, for simplicity) to SmB$_6$, as illustrated in Fig. \ref{tunnelmodel}. First, when a good tunnel barrier is formed like in the case of plasma oxidized SmB$_6$, the single-step elastic tunneling is predominant as depicted in the left panel, resulting in clear features reflecting the DOS of SmB$_6$. On the other hand, if the deposited Al forms an inter-diffusion layer, electrons can not tunnel but diffuse through it, as depicted by the trajectories in the right panel of Fig. \ref{tunnelmodel}. They will lose energy via inelastic scattering events in that region, effectively rendering those features lost in the current-voltage characteristics. If the Al layer is thick enough to leave a part of it unoxidized on top of the inter-diffusion layer, or if the diffusion is largely reduced at low temperature or due to a diffusion barrier, they tunnel into the unoxidized Al layer instead of SmB$_6$. 

A possible solution for both inter-diffusion and under-oxidation is depositing a thinner ($<$ 10 {\AA}) Al layer at liquid nitrogen temperature and thermally oxidizing it completely. A potential issue would be uneven coverage due to the reduced kinetic energy of sputtered Al particles. Other possibilities include reactive sputtering of Al under a small oxygen partial pressure or direct deposition of Al$_2$O$_3$ using RF sputtering. Also, one can think of using different materials for the tunnel barrier since they may not suffer from the issues raised by Al. The extent of diffusion and sticking of sputtered particles on a given substrate (SmB$_6$ in our case) relatively depends on the pair of materials. Although Al has some serious issues when paired with SmB$_6$ as we report here, it is a well-known material to grow on Nb to form an excellent tunnel barrier \cite{Wolf85}.

The success of plasma oxidation of the SmB$_6$ crystal as a means to form a tunnel barrier has to do with the stability of resultant B$_2$O$_3$ with large enough band gap \cite{Li96} as mentioned earlier. Such methods of utilizing self-oxides as a tunnel barrier have been known for some materials in the literature \cite{Wolf85} and was attempted on SmB$_6$ with some success \cite{Amsler98}. In our case, the emergent topological surface states in SmB$_6$ and their robustness against non-magnetic perturbations \cite{KimDJ14} are very beneficial. While it remains to investigate detailed properties of the B$_2$O$_3$ layer formed on SmB$_6$, the cleanliness and reproducibility of the conductance features suggests that it has a high-enough potential barrier sharply interfaced with both SmB$_6$ and Pb, as illustrated in the left panel of Fig. \ref{tunnelmodel}. Here, the majority of electrons undergo a single-step elastic tunneling and, thus, the differential conductance maps out the DOS of SmB$_6$ when the Pb is in the normal state. Also, it is interesting to note that, after the plasma oxidation, the surface states must have moved down to underneath the B$_2$O$_3$ layer, attesting their robustness, thus, topological nature. If they were trivial metallic states at the surface, they would not be able to survive such harsh processes as polishing and plasma oxidation.

As revealed in our conductance spectra and analysis \cite{Park16}, the topological surface states in SmB$_6$ are not intact under the existence of bulk excitations, unlike in the weakly correlated counterparts. Furthermore, the temperature dependence of the conductance, both $G(V)$ and $G(V_b)$, indicates that their spectral density undergoes a rather complex evolution due to the interaction with spin excitons \cite{Kapilevich15}, which must also impact many other physical properties measured. For instance, the temperature dependence of DC resistivity in SmB$_6$ has been interpreted as exhibiting temperature-dependent activation energy gaps \cite{Flachbart01} when analyzed using the conventional Arrhenius expression as for typical semiconductors. We conjecture that such non-trivial temperature dependence might originate from the complex evolution of the surface states rather than the bulk hybridization gap itself. Also, the helical spin texture \cite{Xu14} might be influenced greatly by the same interaction.

Finally, we mention broader implications of our findings on other topological materials in which strong correlations govern their ground states. Their phase diagrams are generally complex due to competing/intertwined orders (e.g., the Doniach phase diagram for Kondo lattices \cite{Doniach77}) and, thus, it is not uncommon that topological phases may emerge in close proximity to quantum critical points. In such cases, there can be strong bulk excitation modes due to critical fluctuations, which will inevitably interact with the topological surface states as we observe in SmB$_6$. Therefore, this possibility should always be taken into account in studying strongly correlated topological materials.

%\vspace{20mm}

\section{Summary}
PTS, being both surface-sensitive and momentum-selective, is a technique suitable for the investigation of electronic properties of the topological surface states in SmB$_6$. The conventional method of depositing and oxidizing a thin Al layer is found to pose several challenges. Not only is the complete oxidation of Al non-trivial but also the inter-diffusion of Al with SmB$_6$ is found to be a fundamental issue, as tested in a few systematic studies with various thickness and deposition schemes for Al. While it is possible to overcome these challenges with a better system, we have successfully identified an alternative method. Here, the key processing step is to form a self-oxide. Namely, the top surface of a SmB$_6$ crystal is plasma-oxidized to form a tunnel barrier, i.e., B$_2$O$_3$ layer as confirmed by an XPS analysis. This method produces high-quality junctions with excellent reproducibility. The tunnel conductance of such junctions reflects spectroscopic properties of the bulk and the surface states as well \cite{Park16}. While the bulk hybridization gap is found to open up below $\sim$50 K with the full gap size of $\sim$21 meV, signatures for the surface states begin to appear at a much lower temperature, $\sim$25 K. At low temperature, the conductance in the low-bias region exhibits double- or single-linear shape for the (001) or (011) surface, respectively. Our analysis shows that this difference between the two surfaces originates from different numbers of distinguishable surface Dirac bands, i.e., two versus one, in good agreement with theoretical calculations \cite{Takimoto11,Lu13,Kim14,Alexandrov13} and quantum oscillation measurements \cite{Li14} as well. Remarkably, the linearity of conductance ends at $\pm$4 mV, well below the gap edges. From an analysis based on the inelastic tunneling model, this premature deviation from the linearity is found to signify the interaction with bulk excitations, spin excitons \cite{Kapilevich15}. Our findings shed new light on the intriguing nature of the topological states in SmB$_6$. Further investigations using even higher-quality tunnel junctions should promise to unveil more details.\\

\section{ACKNOWLEDGMENTS}
The authors acknowledge M. Dzero, L. H. Greene, A. Noddings, and P. Riseborough for fruitful discussions, and C. O. Ascencio and R. Haasch for the help with surface analyses. This material is based upon work supported by the US National Science Foundation (NSF), Division of Materials Research (DMR) under Award 12-06766, through the Materials Research Laboratory at the University of Illinois at Urbana-Champaign. The work done at the University of California-Irvine was supported by the US NSF DMR under Award 08-01253.

%\bibliography{Paper_AC-PCS_RSI_arXiv.bbl}
%\bibliography{PTS_SmB6}
\bibliography{C:/wkpark/PaperNpresentation/Bibliography/TeX/SingleBibForTeX}

\end{document}